 \documentclass[final,5p,times,twocolumn,authoryear]{elsarticle}
\usepackage{amsmath}
\usepackage{amssymb}
\usepackage{graphicx}
\usepackage{color}

\journal{Extreme Mechanics}
\begin{document}
\begin{frontmatter}
\title{Microstructural model for calponin stabilization of crosslinked actin networks}

\author[hlm]{Horacio L\'opez-Men\'endez\corref{cor1}}
\ead{horacio.lopez-menendez@ijm.fr / horacio.lopez.menendez@gmail.com}
\author[jfr]{Jos\'e F\'elix Rodr\'iguez}
\ead{josefelix.rodriguezmatas@polimi.it}
\cortext[cor1]{Principal corresponding author}
\address[hlm]{Cell Adhesion and Mechanics, Institut Jacques Monod (IJM), CNRS UMR 7592 \& Universit\`e Paris Diderot, Paris, France}
\address[jfr]{LaBS, Department of Chemistry, Materials and Chemical Engineering "Giulio Natta", Politecnico di Milano, Piazza Leonardo da Vinci 32, 20133 Milano, Italy}     

\begin{abstract}

The synthetic actin network demands great interest as bio-material due to its soft and wet nature that mimic many biological scaffolding structures. The study of these structures can contribute to a better understanding of this new micro/nano technology and the cytoskeleton like structural building blocks. Inside the cell the actin network is regulated by tens of actin-binding proteins (ABP's), which makes for a highly complex system with several emergent behavior. In particular Calponin is an ABP that was identified as an actin stabiliser, but its mechanism is poorly understood. Recent experiments using an in vitro model system of cross-linked actin with Calponin and large deformation bulk rheology, found that networks with basic calponin exhibited a delayed onset and were able to withstand a higher maximal strain before softening. 

In this work we propose a mathematical model into the framework of nonlinear continuum mechanics to explain the observation of Jansen, where we introduce the hypothesis by which the difference between the two networks, with and without Calponin, is interpreted as a alterations in the pre-strain developed by the network entanglement and the regulation of the crosslinks adhesion energy. The mechanics of the F-actin is modelled using the wormlike chain model for semi-flexible filaments and the gelation process is described as mesoscale dynamics for the crosslinks. The model has been validated with reported experimental results from literature, showing a good agreement between theory and experiments.
 
\end{abstract}

\end{frontmatter}

\section*{Introduction}


Actin network is one of the most relevant structural component inside cytoskeleton of eukaryotic cells, this have a highly dynamic mechanical properties in the cell. Its dynamics is essential to many process, such as cell adhesion, mechanosensing and mechanotransduction, motility, and cell shape among others. At the same time the development of actin based biomimetic systems at micro and nano-scale demand a deep understanding of their mechanical properties at large deformation where the non-linear mechanics effect encode effects that allow the tuning of unexpected effects \citep{Schmoller2010,Schmoller2013,lopez2016}.

In the cell, the dynamics and mechanics of the actin cytoskeleton are regulated by upward sixty known actin-binding proteins (ABPs) defining different levels with emergents behaviour with strong implications in physiological and pathological conditions. In particular the Calponin ABP was discovered in smooth muscle cells and was studied as possible regulator of actomyosin interaction \citep{winder1990}. Non-muscle calponin are known to be involved in actin stabilisation, stress-fibres formation and increase the tensile strength of the tissue under strain among others effect \citep{jensen2012}. Despite the growing evidence on the role performed by Calponin as a structural stabilisers the underlying micro-mechanics still unknown.

To study the mechanical effect developed by Calponin over the actin structure \cite{jensen2014} studied an in-vitro F-actin network crosslinked with Calponin to gain insights about its mechanical properties using large deformation rheology. They found that the networks with Calponin are able to reach a higher failure stress and strain while reducing the pre-strain of the network. Calponin delays the onset of network stiffening, something observed in polymer networks with increased flexibility. They also observed that the micro-structural origin of these behavior was related to the decrease on the persistence length at single filament level.


In order to explain the observed effect reported by \cite{jensen2014} we develop a mathematical model within the framework of continuum mechanics. In this model, we introduce the hypothesis by which the difference between the two networks, with and without Calponin, is interpreted as a alterations in the pre-strain developed by the network entanglement and the regulation of the crosslinks adhesion energy. As a consequence, when the network have a higher internal pre-strain their crosslinks are also near the fluidisation transition.


According to the observed stress-strain relationship at the concentrations of cross-linkers and actin used in the experiment,  it can be assumed continuum strain, with elasticity originating from the entropic nature of the individual polymers \citep{jensen2014, shin2004}. In the first part of this work we describe the mechanics of actin networks with rigid crosslinks, using the wormlike chain model in the form proposed by~\cite{Mackintosh1995} and further developed by \cite{Palmer2008}, whereas the network is described using an homogenized continuum framework based on the eight chain network \citep{Arruda1993,Bertoldi2007,Palmer2008,Brown2009,lopez2016}. Afterwards, we introduce the inelastic effect driven by the crosslinks as alterations in the contour length of the F-actin network. To define the phenomenological law that drives the changes in the contour length, we propose a simple model for the gelation process of the network based on the interactions between the entanglement and adhesion energy of crosslinkers. Finally we compare with experimental measurements performed by \cite{jensen2014}

\section{Mechanics F-actin network with weak crosslinks}\label{network}

Let $\Omega_{0}$ be a fixed reference configuration of the continuos body of interest (assumed to be stress free). We use the notation $\mathbf{\chi}:\Omega_{0}\rightarrow\mathbb{R^{\text{3}}}$ for the deformation, which transforms a typical material point $\mathbf{X}\in\Omega_{0}$ to a position $\mathbf{x}=\mathbf{\chi}(\mathbf{X})\in\Omega$ in the deformed configuration. Further, let  $\mathbf{F}(\mathbf{X})=\frac{\partial\mathbf{\chi}(\mathbf{X})}{\partial\mathbf{X}}$ be the deformation gradient and $J(\mathbf{X})=\det\mathbf{F}(\mathbf{X})>0$ the local volume ratio. Consider a multiplicative split of $\mathbf{F}$ into spherical (dilatational) part, $\mathbf{F}_{V}=(J^{1/3}\mathbf{I})$ and a uni-modular (distortional) part $\bar{\mathbf{F}}$. Note that $\det\bar{\mathbf{F}}=1$. 

\begin{figure}[h!]
	\includegraphics[width=6cm]{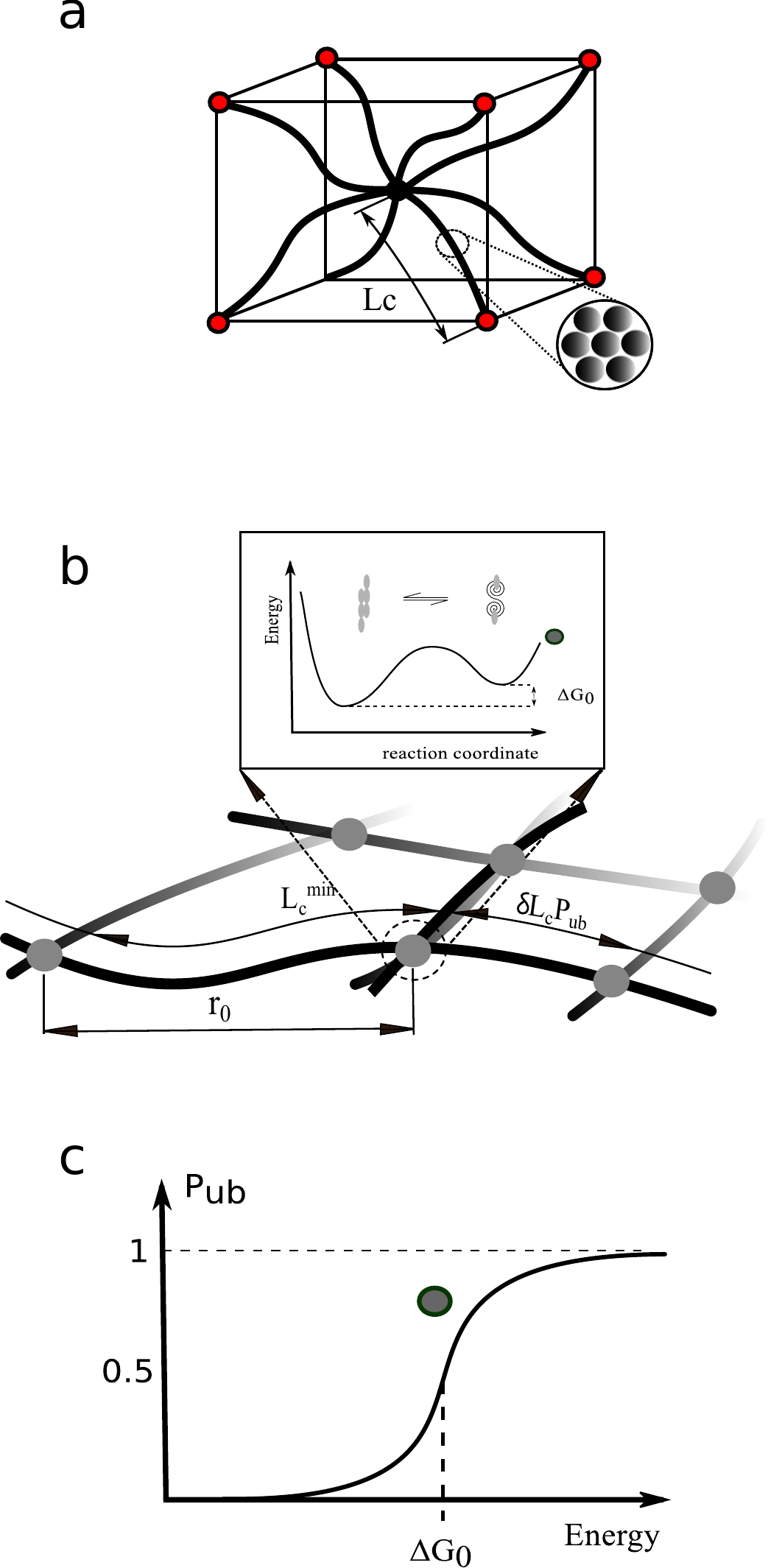}
	\centering
	\caption{Elements of the dynamic F-actin network. $\mathbf{a}$) Eight chains model. $\mathbf{b}$) Semi-flexible bundle filaments in which the contour length is defined as the distance between the crosslinks. $\mathbf{c}$) Unbinding probabiity.}
	\label{dynamic_net}
\end{figure}

We use the right and left Cauchy-Green tensors denoted by $\mathbf{C}$ and $\mathbf{b}$, respectively, and their modified counterparts associated with $\bar{\mathbf{F}}$, $\bar{\mathbf{C}}$ and $\bar{\mathbf{b}}$, respectively. Hence,
\begin{equation}
\mathbf{C}=\mathbf{F}^{T}\mathbf{F}=J^{2/3}\bar{\mathbf{C}},
\end{equation}
\begin{equation}
\bar{\mathbf{C}}=\mathbf{\bar{F}}^{T}\mathbf{\bar{F}},
\end{equation}
\begin{equation}
\mathbf{b}=\mathbf{F}\mathbf{F}^{T}=J^{2/3}\bar{\mathbf{b}},
\end{equation}
\begin{equation}
\bar{\mathbf{b}}=\bar{\mathbf{F}}\bar{\mathbf{F}}^{T}.
\end{equation}

The mechanical behavior of the F-actin cross-linked network is modelled by means of a strain energy function (SEF) based on the wormlike chain model for semi-flexible filaments. This model, proposed firstly by~\cite{Mackintosh1995} and later in a polynomial and homogenised form by~\cite{Palmer2008}, is defined in terms of four physical parameters related to the network architecture and network deformation (see Figure~\ref{dynamic_net}): i) The contour length, $L_{c}$; ii) The persistence length, $l_{p}$; iii) The end-to-end length at zero deformation, $r_{0}$, associated with the network prestress~\citep{Palmer2008}; and iv) The macroscopic network stretch from the condition of zero force, $\lambda:=\sqrt{\bar{I}_1/3}$. 

\begin{eqnarray}
\Psi_{wlc}(\mathbf{C},r_{0},l_{p},L_{c})&=&\psi_{wlc}(\bar{\mathbf{C}},r_{0},l_{p},L_{c})+U(J) \label{eq_W}
\end{eqnarray}
with
\begin{eqnarray}
\psi_{wlc}=\frac{nk_{B}T}{l_{p}}\left[\frac{L_{c}}{4\left(1-\displaystyle{\frac{r}{L_{c}}}\right)^{2}}-l_{p}\log\frac{L_{c}^{2}-2l_{p}L_{c}+2l_{p}r}{r-L_{c}}\right],\label{eq_W1}
\end{eqnarray}

where $r:=r_0\lambda=r_0\sqrt{\bar{I}_1/3}$ is the filament end-to-end distance, $\bar{I}_1$ the first invariant of $\bar{\mathbf{C}}$.
 Using standard procedures from Continuum Mechanics, the Cauchy stress,
$\boldsymbol{\sigma}$, can be derived from direct differentiation
of Eq.~\ref{eq_W} with respect to $\mathbf{C}$ 
\begin{equation}\label{eq_5_1}
\begin{split}\boldsymbol{\sigma} & =\frac{2}{J}\mathbf{F}\frac{\partial\Psi_{wlc}}{\partial\mathbf{C}}\mathbf{F}^{T}\\
 & =\frac{1}{J}\frac{nk_{B}T}{3l_{p}}\frac{r_{0}}{\lambda}\left[\frac{1}{4\left(1-\frac{\lambda r_{0}}{L_{c}}\right)^{2}}\right]\left[\frac{\frac{L_{c}}{l_{p}}-6\left(1-\frac{\lambda r_{0}}{L_{c}}\right)}{\frac{L_{c}}{l_{p}}-2\left(1-\frac{\lambda r_{0}}{L_{c}}\right)}\right]\left(\bar{\mathbf{b}}-\frac{1}{3}\bar{I}_1\mathbf{I}\right)+p\mathbf{I},
\end{split}
\end{equation}
where $p=dU/dJ$ is a constitutive relation for the dilatational part of $\boldsymbol{\sigma}$. For a simple shear test, $\lambda=\sqrt{1+\gamma^{2}/3}$, and the stress-strain relation results
\begin{equation}\label{tau_wlc}
\tau=\frac{nk_{B}T}{3l_{p}}\frac{r_{0}}{\lambda}\left[\frac{1}{4\left(1-\displaystyle{\frac{\lambda r_{0}}{L_{c}}}\right)^{2}}\right]\left[\frac{\displaystyle{\frac{L_{c}}{l_{p}}}-6\left(1-\frac{\lambda r_{0}}{L_{c}}\right)}{\displaystyle{\frac{L_{c}}{l_{p}}}-2\left(1-\frac{\lambda r_{0}}{L_{c}}\right)}\right]\gamma. \\[2pt]
\end{equation}

As mentioned previously, in vivo or in vitro actin networks experience prestress during network formation and remodelling due to the entanglement and the formation and disruption of crosslinks. In order to account for these effects, we introduce a passive prestress, in the network model, through the parameter $r_0$  

\begin{equation}\label{eq_r0}
r_{0}=(1+\epsilon)L_{c}\left(1-\frac{L_{c}}{6l_{p}}\right),
\end{equation}
where $\epsilon>0$ is a dimensionless parameter associated with the passive prestress. 

The network is buildup by the interaction between the actin filaments and the crosslinkers. The nature of this interaction defines the mechanical properties of the structure. If these interactions are stable (for the stress and the time scales of the experiments), they provide a strong gelation process and the meshwork shows a solid-like behaviour under deformation. If on the contrary, the crosslinks are not completely stable, but they are associated with a reaction that can proceed in both directions, folding/unfolding or flexible/rigid states of the crosslink, we then speak of a weak gelation process with the meshwork showing a fluid-like behaviour without manifesting a complete unbinding. Clearly if the level of deformation exerted over the crosslink exceeds a given threshold it will break irreversibly.

These effects are account within the model via the contour length, $L_c$. We propose the following expression for $L_c$
\begin{equation}
L_{c}=L_{c}^{\min}+\delta L_{c}P_{uf},
\end{equation}
where $P_{uf}$ defines the unfolding probability encompassing the states of unfolding or flexible cross-link (see bellow), $L_{c}^{\min}$ represents the contour length when $P_{uf}=0$, and $\delta L_{c}$
represents the average increment of the contour length when the unbinding probability is one. 
Chemical crosslinks can be modelled as a reversible two-state equilibrium process \citep{Brown2009,purohit2011}: 

\begin{equation}
\frac { P_{ uf} }{ P_{ f} } =\exp -\frac {( \Delta G_{ 0 } - w_{ ext} )}{ k_{ B }T },  \label{eq_7}
\end{equation}
where $P_f$ the binding probability encompassing the states folding or rigid cross-link. Since only these two states are possible, then $P_{uf} + P_f = 1$. The two-state model has the folded state as the preferred low free energy equilibrium state at zero force and the unfolded state as the high free energy equilibrium state at zero force. $\Delta G_0$  represents the difference in the free-energy between these states, $w_{ext}$ represents the external mechanical work that induce the deformation of the crosslink, and $k_{B}T$ represents the thermal energy. 

As we are only able to measure macroscopic quantities as stress and strain and we aim to develop a constitutive model in the mesoscale, we propose the next phenomenological expression, using the previous expression as motivation:
\begin{equation}\label{eq_11}
P_{uf}=\frac{1}{1+\exp\left[\kappa\left(\lambda_{0}-\lambda\right)\right]},
\end{equation}
where the main driving force is $\lambda$, the average stretch over the bundle. In order to simplify the mathematical treatment, we consider $\kappa\lambda_0$  to be proportional to the free adhesion energy $\Delta G_0$.  Parameter $\kappa$ gives us an idea of the sharpness of the transition between states and $\lambda_0$ is the strain at which the probability of unbinding transition is 0.5. If $\lambda_{0} << \lambda$  , the network is easy to be remodelled showing a more fluid-like behavior. On the contrary, if $\lambda_{0} >> \lambda$, the crosslink is more stable and the probability of transition is low. Consequently the network behaves as a solid-like structure.

\begin{figure}[h]
	\includegraphics[width=7cm]{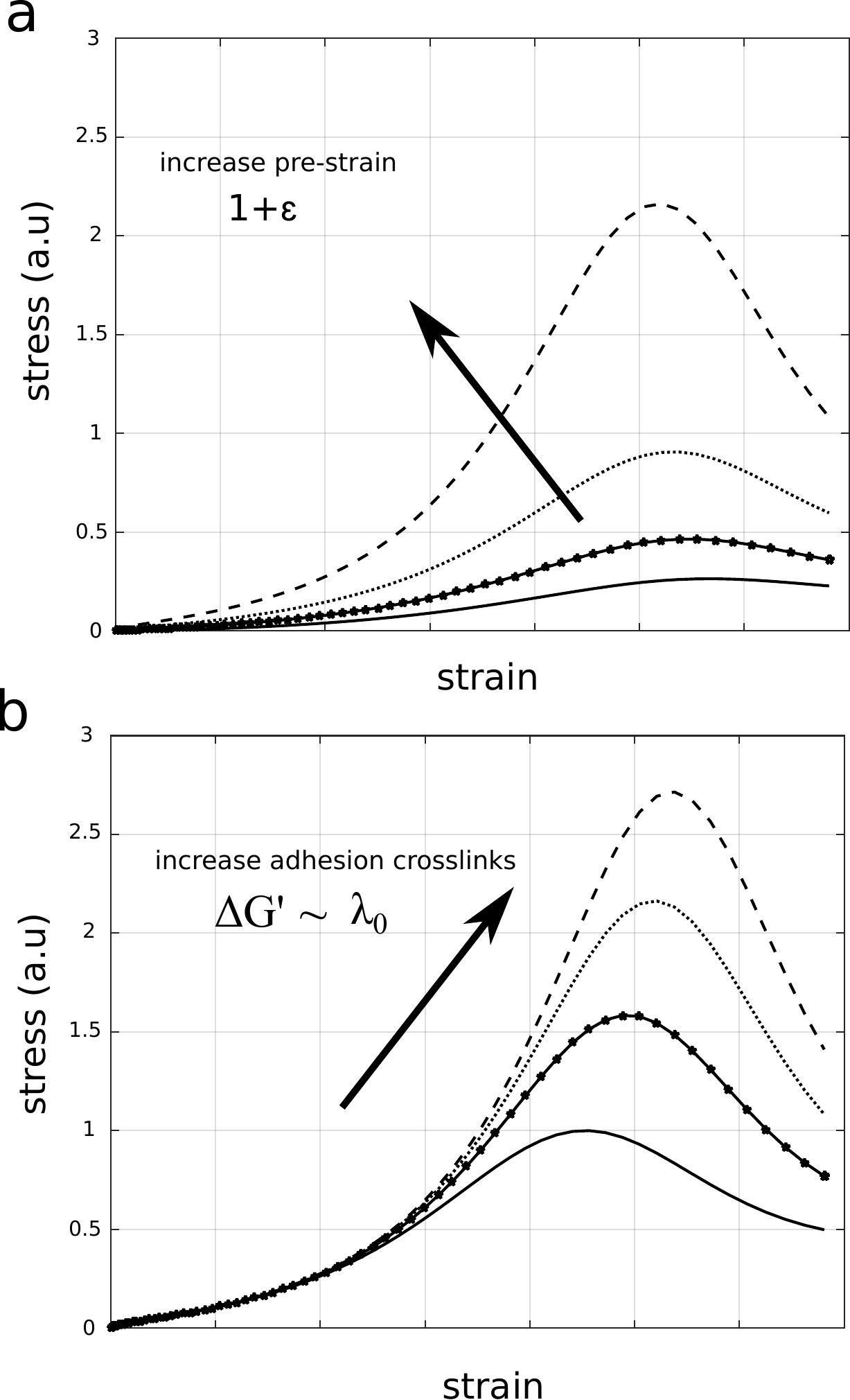}
	\centering
	\caption{(a) Describes the network response for different levels of pre-strain. It shows the increment in the slope for low values of network deformation. (b) Describes the network response for increasing values of $\lambda_0$ showing an extension of the solid-like regime.}
	\label{quant_resp}
\end{figure}

The model of F-actin network with weak crosslinks subjected to simple shear can be summarized as follows: 

\begin{equation}\label{tau_wlc}
\tau=\frac{nk_{B}T}{3l_{p}}\frac{r_{0}}{\lambda}\left[\frac{1}{4\left(1-\displaystyle{\frac{\lambda r_{0}}{L_{c}}}\right)^{2}}\right]\left[\frac{\displaystyle{\frac{L_{c}}{l_{p}}}-6\left(1-\frac{\lambda r_{0}}{L_{c}}\right)}{\displaystyle{\frac{L_{c}}{l_{p}}}-2\left(1-\frac{\lambda r_{0}}{L_{c}}\right)}\right]\gamma, \nonumber
\end{equation}

\begin{equation}
r= (1+\epsilon) L_{c}\left(1-\frac{L_{c}}{6l_{p}}\right)\lambda,     \nonumber
\end{equation}

\begin{equation}
L_{c}=L_{c}^{\min}+\frac{\delta L_{c}}{1+\exp\left[\kappa\left(\lambda_{0}-\lambda\right)\right]}. \nonumber
\end{equation}

In order to describe qualitatively the behaviour of the coupled set of equations under alterations in the pre-strain and the adhesion energy of the crosslinks we evaluate them in the regime of semi-flexible response i.e $L_c \propto l_p$. 
Figure \ref{quant_resp}.a describes the effect of an increment on the pre-strain (1+$\epsilon$) on network response, with the remaining parameters kept constant. As can be observed, as the pre-strain increases, the network stiffness increases and is able to reach a higher level of stress (higher yield point). However, the yielding point (fluidization of the network) occurs earlier reducing the solid-like regime of the network. Figure \ref{quant_resp}.b shows the response of the network for different values of $\lambda_0$. Contrary to the pre-strain, as $\lambda_0$ increases the initial stiffness of the network remains unaltered while the yielding stress and strain increase, extending the solid-like regime. This implies that as $\lambda_0$ increases the crosslinks become more stable. 

\section{Results}

The proposed theory is used to describe the experiments conducted by \cite{jensen2014} on the artificially reconstituted F-actin networks crosslinked with filamin, where the network has an actin decorated with and without Calponin. We evaluate the proposed model for the set of parameters shown in Table~\ref{table1} identified by means of nonlinear least-square fit to experiments of monotonic large deformation stress-strain shear tests from~ \cite{jensen2014}. 

The parameters of the model shown in Table~\ref{table1} can be divided in two types: (i) Rigid-wormlike chain parameters $L_{c}^{\min}$, $l_{p}$, $\delta L_{c}$ and $\epsilon$ which are of the order of magnitude of the values used to describe in experiments of in-vitro F-actin networks and to keep on the regime the regime of semi-flexible entropic elasticity.  (ii) Parameters associated with the dynamics of the crosslinks $\kappa$ and $\lambda_{0}$. These parameters encode the transitions to induce fluidisation of the network and represent an indirect measure of the adhesion force of crosslinks.  While $\lambda_0$ describes the transition point in the contour length of the network meshsize (average distance between crosslinks), $\kappa$ describes the sharpness of this transition. These values are phenomenological and were identified in order to fit the experimental data.
 
 \begin{figure}[h!]
	\includegraphics[width=7.5cm]{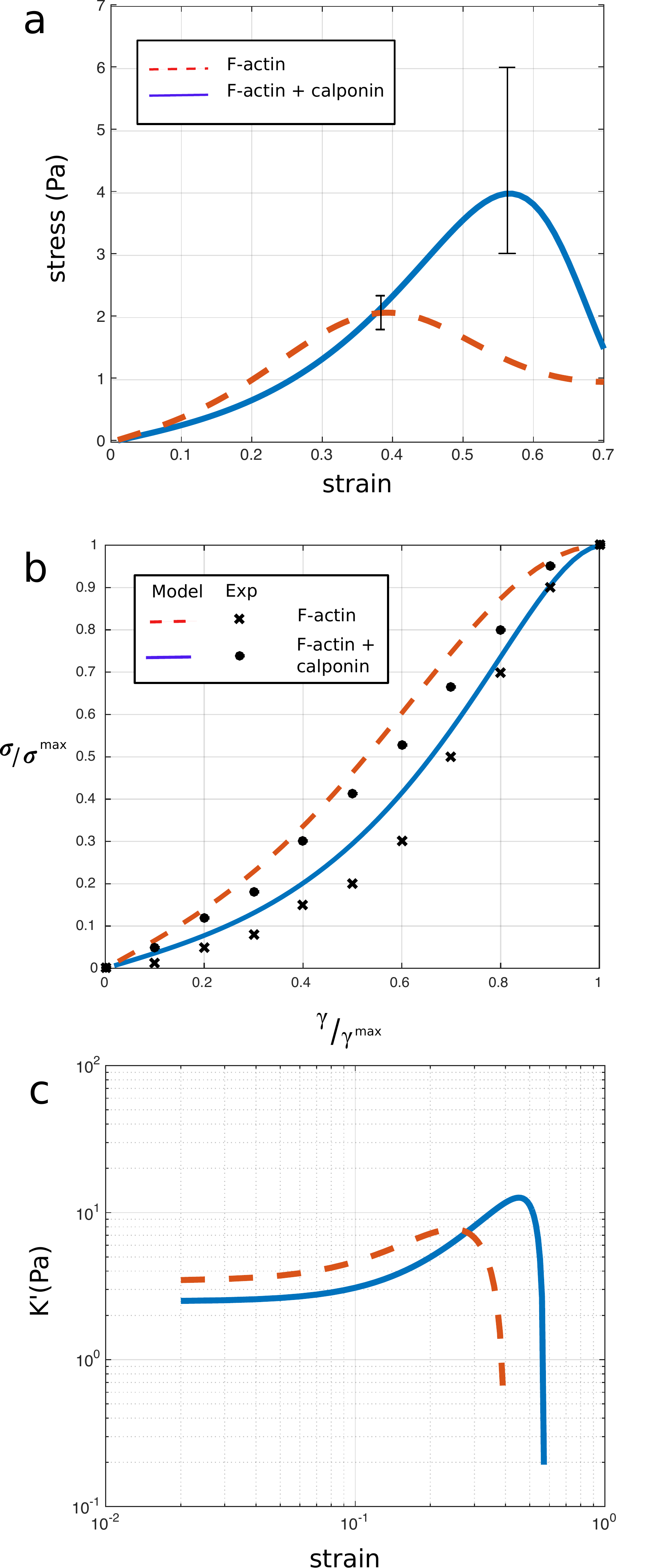}
	\centering
	\caption{(a) Model simulation for the bulk rheological stress-strain curves with and without calponin showing the onset in the transition from strain-stiffening to strain-softening. (b) Normalised stress-strain relation respect the critical values and comparison between theoretical model and experimental results. (c) Differential elastic modulus $K\approx\frac{d\sigma}{d\gamma}$. }
	\label{res}
\end{figure}

We compared the results obtained by the model with the average curve of monotonic simple shear at a constant shear strain rate. The network stiffening response begins at low levels of strain and continues almost linearly until reaching a maximum critical shear stress.  After that point, the response change towards a regime of stress-strain softening where the stress decreases slowly towards zero as the shear strain increases, and the structure flows.

For low levels of Calponin the networks present a higher level of pre-strain as noted by the higher slope at low level of deformation, which can also can be observed in the Figure \ref{res}.c as a larger value of $K\approx\frac{d\sigma}{d\gamma}$. For low levels of Calponin strain stiffening occurs until $\gamma_{max}\approx 0.4$, with a maximum stress $\tau_{max}\approx 2 Pa$ (see Figure \ref{res}.a). On the contrary, for higher levels of Calponin, the response shows a lower level of pre-strain (lower initial stiffness), but the yielding point occurs at a larger deformation $\gamma_{max}\approx 0.55$ and higher stress $\tau_{max}\approx 4 Pa$ (experimental values between 3 and 6 $Pa$) as shown in  Figure~\ref{res}.a, indicating that Calponin extends the solid-like behaviour of the network. This is a remarkable result with respect to pure F-actine networks for which a lower level of pres-train increases the solid-like regime, but reduces the yield stress significantly. Figure \ref{res}.b shows the scaled stress-strain curve with respect to the yield point. This figure demonstrates the ability of the model to describe the nonlinear effects.

\section{Discussion and Conclusions}

In order to explain the effects of Calponin in stabilisation of the crosslinked F-actin networks 
our propose that the observed effect  which promotes by modification in the flexural rigidity, and as a consequence the persistence length, at single filament will integrate at network scale as a change in the internal pre-stress and the adhesion energy over the crosslinks.

\begin{figure}[h]
	\includegraphics[width=8cm]{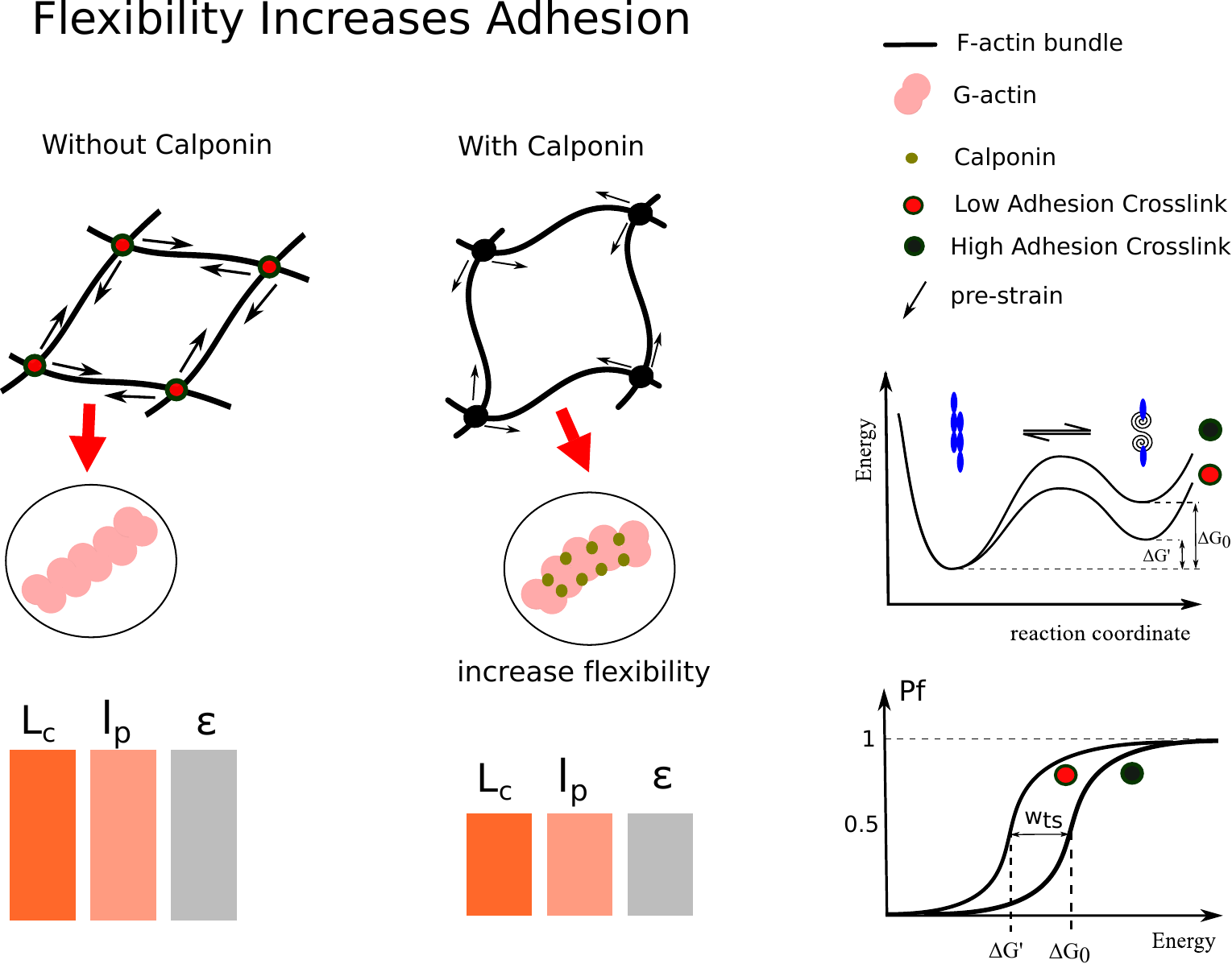}
	\centering
	\caption{Network scale effects with and without Calponin.The role of Calponin is to increase the flexibility between the bonds of G-actin that connect each other to create the actin filament.
Without Calponin the persistence length increase promoting the increment in pre-strain over bundle and crosslinkers which also increment the contour length, shifting the network behaviour towards the fluid-like regime. With Calponin the increased flexibility of the filaments creates a network with lower pre-stress and higher adhesion energy in the crosslinks, reducing the contour length and extending the solid-like regime.
	}
	\label{calp_eff}
\end{figure}

We condense these observations in the Figure \ref{calp_eff}. During the gelation process, the network entanglement induces pre-stress across the network which is propagated through the bundles to the chemical crosslinks \citep{Lieleg2009,Lieleg2011}. The trapped stress in the structure is compensated by the deformation of the bundle and the chemical crosslinks. As a consequence, it is potentially able to induce conformational changes over the crosslink structure, as was described by \cite{Golji2009}.
The red and black dots in Figure~\ref{calp_eff} describe the effect of the pre-stress on the energy landscape of the chemical crosslinks. Qualitatively, results easy to see that the energy gap is lower under the presence of pre-stress on the crosslink, where the adhesion energy increases, changing from $\Delta G'$ to $\Delta G_0$, with $\Delta G_0 > \Delta G'$. This can be understood as a combined action of two mechanical regulation pathways over the bonds reaction, where: $i)$ $w_{ext}$ represents the mechanical work induced by the macroscopic deformation that propagates through the network down to the crosslink. $ii)$ $w_{ts}$ represents the mechanical work introduced during the entanglement also deforms the crosslink structure.  On one side, without Calponin the bundles are less flexible, and therefore network deformation will deform more the crosslink, which in terms of the energy landscape, it appears as tilted down (se Figure~\ref{calp_eff}), and therefore facilitates the conformational change from the folded configuration to the unfolded configuration. Under the perspective of the proposed model, it implies that $\lambda_0$ is shifted to the left, and the pre-strain $(1+\epsilon)$ is higher. Therefore, the network is closer to the fluidisation transition and the maximum stress (yield stress) is reduced.
On the contrary, if the concentration of calponin is higher, the bundle flexibility increases and persistence length decrease implying that the effects associated with trapped stress will be more concentrated on the bundle deformation than on the crosslinks. In consequence the adhesion energy of the crosslink is higher (more difficult to jump from the folded to the unfolded configuration) and the level of shear deformation that the network is able to achieve before the solid-fluid transition results also is also higher. In summary, the proposed model provides arguments to describe the changes observed in the flexibility of the actin bundles encoded the effects at network scale that drives the increment of adhesion and the whole stabilisation of the structure. This will help to gain better understanding in the complex mechanics behind of the cytoskeleton-like structural building blocks and synthetic bio-structures.


%
%
%
%
%
%
%
%


\begin{table}
\centering
\begin{tabular}{lll}
\hline

$n$&Density of actin filaments& $1.e22$ [m$^{-3}$]\tabularnewline
$k_BT$& Thermal energy & $4.1$ [$p$N$n$m]\tabularnewline
$L_{c}^{min}$ & Contour length & $13$ [$\mu$m]\tabularnewline
$\delta L_{c}$ & Contour length & $2L_{c}^{min}$ \tabularnewline
\textbf{Low  [Calponin]} \tabularnewline
$l_{p}$ & Persistence length & $10$ [$\mu$m]\tabularnewline
$\kappa$ & Nondim. crosslinks stiffness & $60$\tabularnewline
$\lambda_{0}$ & Characteristic stretch  & $1.04$\tabularnewline
$1+\epsilon$ & Bundle prestrain & $1.01$\tabularnewline
\textbf{High [Calponin]} \tabularnewline
$l_{p}$ & Persistence length & $6$ [$\mu$m]\tabularnewline
$\kappa$ & Nondim. crosslinks stiffness & $75$\tabularnewline
$\lambda_{0}$ & Characteristic stretch  & $1.09$\tabularnewline
$1+\epsilon$ & Bundle prestrain & $1.02$\tabularnewline

\hline
\end{tabular}
\caption{Model parameters to fit the experiments of \cite{jensen2014} }\label{tabR1}
\label{table1}
\end{table}

\section*{References}


\begin{thebibliography}{16}
\providecommand{\natexlab}[1]{#1}
\providecommand{\url}[1]{\texttt{#1}}
\expandafter\ifx\csname urlstyle\endcsname\relax
  \providecommand{\doi}[1]{doi: #1}\else
  \providecommand{\doi}{doi: \begingroup \urlstyle{rm}\Url}\fi

\bibitem[Arruda and Boyce(1993)]{Arruda1993}
E.~Arruda and M.~Boyce.
\newblock {A three-dimensional constitutive model for the large stretch
  behaviour of rubber elastic materials}.
\newblock \emph{Journal of the Mechanics and Physics of Solids}, 41:\penalty0
  389--412, 1993.

\bibitem[Bertoldi and Boyce(2007)]{Bertoldi2007}
K.~Bertoldi and M.~Boyce.
\newblock {Mechanics of the hysteretic large strain behavior of mussel byssus
  threads}.
\newblock \emph{Journal of Materials Science}, 42\penalty0 (21):\penalty0
  8943--8956, 2007.

\bibitem[Brown et~al.(2009)Brown, Litvinov, Discher, Purohit, and
  Weisel]{Brown2009}
A.~Brown, R.~Litvinov, D.~Discher, P.~Purohit, and J.~Weisel.
\newblock {Multiscale mechanics of fibrin polymer: gel stretching with protein
  unfolding and loss of water.}
\newblock \emph{Science}, 325\penalty0 (5941):\penalty0 741--4, 2009.

\bibitem[Golji et~al.(2009)Golji, Collins, and Mofrad]{Golji2009}
J.~Golji, R.~Collins, and M.~Mofrad.
\newblock Molecular mechanics of the $\alpha$-actinin rod domain: Bending,
  torsional, and extensional behavior.
\newblock \emph{PLoS computational biolog}, 5, 2009.

\bibitem[Jensen et~al.(2012)Jensen, Watt, Hodgkinson, Gallant, Appel,
  El-Mezgueldi, Angelini, Morgan, Lehman, and Moore]{jensen2012}
M.~Jensen, J.~Watt, J.~Hodgkinson, C.~Gallant, S.~Appel, M.~El-Mezgueldi,
  T.~Angelini, K.~Morgan, W.~Lehman, and J.~Moore.
\newblock Effects of basic calponin on the flexural mechanics and stability of
  f-actin.
\newblock \emph{Cytoskeleton}, 69\penalty0 (1):\penalty0 49--58, 2012.

\bibitem[Jensen et~al.(2014)Jensen, Morris, Gallant, Morgan, Weitz, and
  Moore]{jensen2014}
M.~Jensen, E.~Morris, C.~Gallant, K.~Morgan, D.~Weitz, and J.~Moore.
\newblock Mechanism of calponin stabilization of cross-linked actin networks.
\newblock \emph{Biophysical journal}, 106\penalty0 (4):\penalty0 793--800,
  2014.

\bibitem[Lieleg et~al.(2009)Lieleg, Schmoller, Claessens, and
  Bausch]{Lieleg2009}
O.~Lieleg, K.~Schmoller, M.~Claessens, and A.~Bausch.
\newblock {Cytoskeletal polymer networks: viscoelastic properties are
  determined by the microscopic interaction potential of cross-links.}
\newblock \emph{Biophysical journal}, 96\penalty0 (11):\penalty0 4725--32,
  2009.

\bibitem[Lieleg et~al.(2011)Lieleg, Kayser, Brambilla, Cipelletti, and
  Bausch]{Lieleg2011}
O.~Lieleg, J.~Kayser, G.~Brambilla, L.~Cipelletti, and A.~Bausch.
\newblock {Slow dynamics and internal stress relaxation in bundled cytoskeletal
  networks}.
\newblock \emph{Nature Materials}, 10\penalty0 (3):\penalty0 236--242, 2011.

\bibitem[L{\'o}pez-Men{\'e}ndez and Rodr{\'\i}guez(2016)]{lopez2016}
H.~L{\'o}pez-Men{\'e}ndez and J.~F. Rodr{\'\i}guez.
\newblock Microstructural model for cyclic hardening in f-actin networks
  crosslinked by $\alpha$-actinin.
\newblock \emph{Journal of the Mechanics and Physics of Solids}, 91:\penalty0
  28--39, 2016.

\bibitem[Mackintosh et~al.(1995)Mackintosh, Kas, and Janmey]{Mackintosh1995}
F.~Mackintosh, J.~Kas, and P.~Janmey.
\newblock {Elasticity of semiflexible biopolymer networks}.
\newblock \emph{Physical Review Letters}, 75:\penalty0 4425, 1995.

\bibitem[Palmer and Boyce(2008)]{Palmer2008}
J.~Palmer and M.~Boyce.
\newblock {Constitutive modeling of the stress-strain behavior of F-actin
  filament networks.}
\newblock \emph{Acta biomaterialia}, 4\penalty0 (3):\penalty0 597--612, 2008.

\bibitem[Purohit et~al.(2011)Purohit, Litvinov, Brown, Discher, and
  Weisel]{purohit2011}
P~Purohit, R~Litvinov, A~Brown, D~Discher, and J~Weisel.
\newblock Protein unfolding accounts for the unusual mechanical behavior of
  fibrin networks.
\newblock \emph{Acta biomaterialia}, 7\penalty0 (6):\penalty0 2374--2383, 2011.

\bibitem[Schmoller and Bausch(2013)]{Schmoller2013}
K.~Schmoller and A.~Bausch.
\newblock Similar nonlinear mechanical responses in hard and soft materials.
\newblock \emph{Nature materials}, 12\penalty0 (4):\penalty0 278--281, 2013.

\bibitem[Schmoller et~al.(2010)Schmoller, Fernandez, Arevalo, Blair, and
  Bausch]{Schmoller2010}
K.~Schmoller, P.~Fernandez, R.~Arevalo, D.~Blair, and A.~Bausch.
\newblock {Cyclic hardening in bundled actin networks}.
\newblock \emph{Nature Communications}, 1:\penalty0 134, 2010.

\bibitem[Shin et~al.(2004)Shin, Gardel, Mahadevan, Matsudaira, and
  Weitz]{shin2004}
JH~Shin, ML~Gardel, L~Mahadevan, P~Matsudaira, and DA~Weitz.
\newblock Relating microstructure to rheology of a bundled and cross-linked
  f-actin network in vitro.
\newblock \emph{Proceedings of the National Academy of Sciences of the United
  States of America}, 101\penalty0 (26):\penalty0 9636--9641, 2004.

\bibitem[Winder and Walsh(1990)]{winder1990}
SJ~Winder and MP~Walsh.
\newblock Smooth muscle calponin. inhibition of actomyosin mgatpase and
  regulation by phosphorylation.
\newblock \emph{Journal of Biological Chemistry}, 265\penalty0 (17):\penalty0
  10148--10155, 1990.

\end{thebibliography}


\end{document}